\documentstyle[emulateapj5,epsf]{aastex}
\def \farcs{\hbox{$.\!\!^{\prime\prime}$}}

\lefthead{Hoekstra et al.}
\righthead{Weak lensing by large scale structure}

\begin{document}


\title{A measurement of weak lensing by large scale structure in RCS fields}

\author{Henk~Hoekstra$^{1,2,3}$, Howard~K.C.~Yee$^{2,3,4}$, 
	Michael D.~Gladders$^{2,3,4}$, L.~Felipe~Barrientos$^{4,5}$,
	Patrick B.~Hall$^{2,3,4,5,6}$, and Leopoldo~Infante$^{4,5}$}

\begin{abstract} 

We have analysed $\sim 24$ square degrees of $R_C$-band imaging data
from the Red-Sequence Cluster Survey (RCS), and measured the excess
correlations between galaxy ellipticities on scales ranging from 1 to
30 arcminutes. We have used data from two different telescopes: $\sim
16.4$ square degrees of CFHT data and $\sim 7.6$ square degrees of CTIO
4-meter data, distributed over 13 widely separated patches. For the first
time, a direct comparison can be made of the lensing signal
measured using different instruments, which provides an important test
of the weak lensing analysis itself. The measurements obtained from
the two telescopes agree well. For the lensing analysis we use
galaxies down to a limiting magnitude of $R_C=24$, for which the
redshift distribution is known relatively well. This allows us to
constrain some cosmological parameters. For the currently favored
$\Lambda$CDM model $(\Omega_m=0.3,~ \Omega_\Lambda=0.7,~\Gamma=0.21)$
we obtain $\sigma_8=0.81^{+0.14}_{-0.19}$ (95\% confidence), in agreement
with the results from Van Waerbeke et al. (2001) which used fainter
galaxies (and consequently higher redshift galaxies). The good
agreement between these two very different weak lensing studies
demonstrates that weak lensing is a useful tool in observational
cosmology.

\end{abstract}


\section{Introduction}

Weak gravitational lensing has proven to be a powerful tool for
studies of the mass distribution in rich clusters of galaxies (for a
review see Mellier 1999). Since the pioneering work by Tyson, Wenk, \&
Valdes (1990) much progress has been made, and nowadays the weak
lensing signals induced by clusters of galaxies at intermediate
redshifts can be measured without much difficulty.

The development of advanced techniques to correct for the various
observational distortions, such as the anisotropy of the point spread
function (PSF), the circularization by the PSF, and the camera
induced distortion, has been a crucial step, resulting in well calibrated
signals (e.g., Kaiser, Squires, \& Broadhurst 1995; Luppino \& Kaiser
1997; Hoekstra et al. 1998; Kuijken 1999; Refregier 2001).  Another
important development in recent years is the advent of mosaic CCD
cameras, which enable us to quickly image large portions of the sky.

These advances have made it possible to pursue one of the most
difficult measurements in the field of weak lensing: the measurement
of the coherent distortions of the images of faint galaxies caused by
lensing by intervening large scale structure, the so-called `cosmic
shear'.  The analysis of this lensing signal provides an important
direct

\vbox{
\vspace{0.3cm}
\footnotesize
\noindent 
$^1$~CITA, University of Toronto, Toronto, Ontario M5S 3H8, Canada\\
$^2$~Department of Astronomy and Astrophysics, University of Toronto, 
Toronto, Ontario M5S 3H8, Canada\\
$^{3}$~Visiting Astronomer, Canada-France-Hawaii Telescope, which is
operated by the National Research Council of Canada, Le Centre
National de Recherche Scientifique, and the University of Hawaii\\
$^{4}$~Guest observer at the Cerro Tololo Inter-American Observatory
(CTIO), a division of the National Optical Astronomy Observatories,
which is operated by the Association of Universities for Research in
Astronomy, Inc., under cooperative agreement with the National Science
Foundation\\ 
$^{5}$~Universidad Cat{\'o}lica de Chile, Depto. Astronom{\'\i}a y Astrof
{\'\i}sica, Avda. Vicu{\~n}a Mackenna 4860, Casilla 306, Santiago 22, Chile\\
$^{6}$~Princeton University Observatory, Princeton, NJ 98544-1001
}

\noindent measurement of the statistical properties of the large scale
mass distribution (e.g., Blandford et al 1991; Kaiser 1992;
Bernardeau, van Waerbeke, \& Mellier 1997; Schneider et al. 1998).

Compared to many other methods that are used to constrain cosmological
parameters, weak lensing has the advantage that no assumptions about
the light distribution are required. However, weak lensing in itself
cannot constrain all the parameters, because of degeneracies between
them. Better constraints can be obtained by comparing weak lensing
studies that probe different redshifts (e.g., Hu 1999) or when these
data are combined with measurements of the fluctuations of the cosmic
microwave background (e.g., Hu \& Tegmark 1999).

By now, several groups have reported the detection of an excess
correlation between galaxy ellipticities, and have argued that this
signal is caused by lensing by large scale structure (e.g., Bacon et
al. 2000; Kaiser, Wilson, \& Luppino 2000; van Waerbeke et al. 2000;
Wittman et al. 2000; Maoli et al. 2001; van Waerbeke et al. 2001).
Maoli et al. (2001) combined their own results with published
measurements in an attempt to obtain constraints on $\sigma_8$, the
normalisation of the power spectrum, and $\Omega_m$, the matter
density of the universe. They found good agreement with studies of
cluster abundances. However, the data set studied by Maoli et
al. (2001) is very inhomogeneous, which limits the accuracy
of such a direct comparison.

After these initial detections, which demonstrated the feasibility of
the method, the obvious next step is to obtain large uniform data
sets. The first results from such a survey were presented by van
Waerbeke et al. (2001), who measured a highly significant lensing
signal from 6.5 square degrees of deep imaging data.

In this paper we present the results from our analysis of $\sim 24$
deg$^2$ of $R_C$-band data from the Red-Sequence Cluster Survey
(RCS) (e.g., Gladders \& Yee 2000), which is a 100 deg$^2$ galaxy
cluster survey designed to provide a large sample of optically
selected clusters of galaxies with redshifts $0.1<z<1.4$. The data are
also useful for a range of lensing studies. For example, Gladders, Yee
\& Ellingson (2001) presented the first results for one of the strong
lensing clusters discovered in the survey, for which follow-up
observations are underway.

The weak lensing applications are numerous. The survey imaging data
are relatively shallow compared to what is common in weak lensing
studies, and as a result the statistical uncertainty in the
measurements of individual structures (such as clusters of galaxies)
is large. However, thanks to the large survey area, many such structures
can be detected, and by stacking the signals, one can study their
ensemble averaged mass distribution (e.g., Hoekstra et al. 2001). In
addition, follow-up observations will provide detailed information on
individual systems.

In this paper we concentrate on the measurement of the weak lensing
signal induced by large scale structure (cosmic shear).  A study of
galaxy biasing, based on some of these data, is presented in Hoekstra,
Yee, \& Gladders (2001), and a study of the properties of galaxy halos
is currently underway.

Compared to other cosmic shear studies, the RCS data are shallow, and
consequently the signal at a given scale is much lower, as is the
signal-to-noise ratio. However, measuring the weak lensing signal from
a shallow survey also has several advantages. Down to a limiting
magnitude of $R_C\sim 24$ star-galaxy separation works well. In deeper
surveys many sources have sizes comparable to the size of the PSF, and
applying size cuts may change the redshift distribution of the sources
in a systematic way. In addition, down to $R_C\sim 24$ the redshift
distribution of the sources is fairly well determined. In order to
relate the observed cosmic shear signal to cosmological parameters, a
good understanding of the source redshift distribution is crucial.

One worry is the effect of intrinsic alignments of the source
galaxies, which introduces an additional signal (e.g., Heavens et
al. 2000; Catelan et al 2001; Crittenden et al. 2001; Mackey et
al. 2001). The amplitude of the effect is not well determined, but it
is clear that it becomes more important for shallower
surveys. However, the predictions indicate that for a median redshift
of $z=0.5$ (which is similar to our sample of source galaxies) the
signal caused by intrinsic alignments is still small compared to the
lensing signal (e.g., Makey et al. 2001), and we will ignore the
effect in this paper.

In \S2 we describe the Red-Sequence Cluster Survey (RCS) from
which we have used the $R_C$-band data for the analysis presented
here.  \S3 deals with the analysis of the data, as well as the
corrections for systematic distortions, such as PSF anisotropy and the
distortion by the camera. In \S4 we discuss the expected signal
from weak lensing by large scale structure. The results of the
analysis are presented in \S5.

\section{Data}

\subsection{The Red-Sequence Cluster Survey}

The Red-Sequence Cluster Survey\footnote{\tt
http://www.astro.utoronto.ca/${\tilde{\ }\!}$gladders/RCS} (RCS) is a
galaxy cluster survey designed to provide a large sample of optically
selected clusters of galaxies with redshifts $0.1<z<1.4$. The planned
survey will cover 100 square degrees in both $R_C$ and $z'$, and
consists of 22 widely separated patches of $\sim 2.1\times 2.3$
degrees.  The northern half of the survey is observed using the CFH12k
camera on the CFHT, and the data from the southern half are obtained
using the Mosaic II camera on the CTIO 4m telescope. The patches are
imaged down to a $5\sigma$ point source depth of 25.2 magnitudes in
the $R_C$-band, and $23.6$ magnitudes in the $z'$ filter.

For the weak lensing analysis we use only the $R_C$ band data, as
these provide a sufficiently high number density of sources to warrant
an accurate measurement of the lensing signal. We present the results
based on $\sim 16.4$ deg$^2$ of CFHT data and $\sim 7.6$ deg$^2$ of
CTIO data.  In this paper we use a subset of the RCS and the data for
each patch are not contiguous. Thus the largest scale we consider here
is that of one pointing of the CFH12k or Mosaic~II camera.

We have used data from all 10 patches that have been observed using
the CFHT, resulting in a total of 53 pointings. The integration times
are 900s per pointing. In addition we have used three patches
(resulting in an additional 23 pointings) from the first CTIO run,
which have integration times of 1200s. Some details about the
observations are listed in Table~\ref{observations}.

\subsection{Data reduction}

Given the large amount of data collected in the survey, special
attention was paid during the survey design on how to handle the data
flow. To simplify the construction of the science images the data were
acquired without dithering. Although the gaps between the chips,
cosmetic defects, and cosmic rays result in a minor loss in area, the
advantage in handling the data flow is tremendous. The loss of area
does not affect the result of the weak lensing analysis, and cosmic
rays are easily removed from the galaxy catalogs.

The individual chips from the mosaic imagers are de-biased, and
flat-fielded using standard techniques. The images are used for the
object analysis, which is described below. A detailed discussion of
the reduction pipeline is presented in Gladders \& Yee (in
preparation).

\section{Object analysis}

Our weak lensing analysis technique is based on that developed by
Kaiser, Squires \& Broadhurst (1995) and Luppino \& Kaiser (1997),
with a number of modifications which are described in Hoekstra et
al. (1998) and Hoekstra et al. (2000). 

This correction scheme assumes that one can model the PSF as
an isotropic function, convolved with a compact, anisotropic
kernel. The method does not make any assumptions about the
profile of the PSF or the galaxy, as these parameters are
measured from the actual data.

In real data, the PSF is likely to be more complex,
and the assumption stated above is not valid. However, as
shown by Hoekstra et al. (1998) matching the measurement of
the PSF parameters to the size of the galaxy results in
accurate corrections for PSFs with varying ellipticity as
a function of radius (as is the case for the WFPC2 PSF).

\vbox{
\begin{center}
\begin{small}
\begin{tabular}{lccc|ccc}
\hline
\hline
patch	& pointing & seeing & run & pointing 	& seeing & run\\
	&	& ['']	 &     &   	& ['']	& \\
\hline
(a) CFHT &	&	&    & 	     &	     &	\\
\hline
0223    & A2    & 0.72  & 2  & B2    & 0.78  & 2  \\
        & A3    & 0.92  & 2  & B3    & 0.79  & 2  \\
        & A4    & 0.77  & 2  & B4    & 0.64  & 2  \\
        & A5    & 0.67  & 2  & B5    & 0.64  & 2  \\
0349    & A1    & 0.69  & 2  & C1    & 0.63  & 2  \\
        & A2    & 0.79  & 2  & C2    & 0.59  & 2  \\
        & A3    & 0.87  & 2  &       &       &    \\
0920    & A2    & 0.73  & 2  & C2    & 0.60  & 1a \\
        & B1    & 0.70  & 2  & C3    & 0.69  & 1a \\
        & B2    & 0.75  & 2  &       &       &    \\
1120    & A3    & 0.80  & 2  & B3    & 0.67  & 2  \\
        & A4    & 0.82  & 2  & B4    & 0.69  & 2  \\
        & B2    & 0.76  & 2  &       &       &    \\
1326    & A3    & 0.74  & 2  & C2    & 0.55  & 1a \\
        & A5    & 0.79  & 2  & C3    & 0.58  & 1a \\
        & C1    & 0.58  & 1a &       &       &    \\
1417    & B2    & 0.61  & 1a & B4    & 0.52  & 1a \\
        & B3    & 0.61  & 1a & B5    & 0.59  & 1a \\
1447    & A2    & 0.79  & 3  & B2    & 0.66  & 2  \\
        & A3    & 0.74  & 3  & B4    & 0.73  & 3  \\
        & B1    & 0.63  & 3  &       &       &    \\
1614    & A1    & 0.59  & 1a & B2    & 0.56  & 1a \\
        & A5    & 0.50  & 1a & B3    & 0.57  & 1a \\
        & B1    & 0.56  & 1a &       &       &    \\
2148    & B2    & 0.63  & 1b & C1    & 0.89  & 1b \\
        & B3    & 0.65  & 1b & C2    & 0.82  & 1b \\
        & B4    & 0.69  & 1b &       &       &    \\
2316    & A1    & 0.69  & 3  & B3    & 0.72  & 3  \\
        & A2    & 0.66  & 3  & B5    & 0.74  & 3  \\
        & A3    & 0.66  & 3  &       &       &    \\
\hline
(b) CTIO	&	&	&   &		&	&   \\
\hline
0333	& A3	& 0.97	& 1 & B4	& 0.82	& 1 \\
	& A4	& 0.98	& 1 & C3	& 1.03	& 1 \\
	& B3	& 0.88	& 1 & C4	& 1.12	& 1 \\
0438	& A1	& 0.80	& 1 & B3	& 0.89	& 1 \\
	& A2	& 0.84	& 1 & B4	& 0.89	& 1 \\
	& A3	& 0.87	& 1 & C3	& 0.94	& 1 \\
	& A4	& 0.94	& 1 & C4	& 0.93	& 1 \\
1102	& A1	& 0.68	& 1 & B4	& 0.77	& 1 \\
	& A2	& 0.75	& 1 & C2	& 0.81	& 1 \\
	& A3	& 0.76	& 1 & C3	& 0.86	& 1 \\
	& A4	& 0.82	& 1 & C4	& 0.79	& 1 \\
	& B3	& 0.81	& 1 &		&	&   \\
\hline
\hline
\end{tabular}
\tabcaption{\footnotesize (a) Some information for the 53 CFHT pointings
used in this analysis. Ten widely separated patches have been
observed. Typically 5 pointings per patch were used, except for the
1417 (4 pointings) and the 0223 patch (8 pointings); (b) Information
for the 23 CTIO pointings.  We used all the data obtained during run~1,
and as a result the number of pointings per patch varies. The seeing
was determined for both telescopes using the sizes of stars on chip~3.
\label{observations}}
\end{small}
\end{center}}

In addition, the accuracy of this method has been studied in great
detail (e.g., Erben et al. 2001; Bacon et al. 2001) and the results
demonstrate that it works well down to the required accuracy for
current data sets.  Hence, the correction scheme is accurate even if
the PSF does not satisfy the assumption made in the
derivation. However, this can be understood easily, because any
residuals induced by higher order moments of the PSF are supressed as
one averages the shapes of galaxies which have random orientations
with respect to the PSF.

We analyse the chips of each pointing separately. After the catalogs
have been corrected for the various observational effects, they are
combined into a master catalog which covers the observed field (for
each pointing). 

The first step in the analysis is to detect the faint galaxy
images, for which we used the hierachical peak finding algorithm
from Kaiser et al. (1995). We select objects which are detected
with a significance greater than $5\sigma$ over the local sky. 

We use single exposures for our analysis, and consequently cosmic
rays have not been removed. However, cosmic rays are readily eliminated
from the photometric catalogs: small, but very significant objects 
are likely to be cosmic rays, or artefacts from the CCD. The
peak finder gives fair estimates of the object size, and we remove
all objects smaller than the size of the PSF. 

The objects in this cleaned catalog are then analysed, which yield
estimates for the size, apparent magnitude, and shape parameters
(polarization and polarizabilities). The objects in this catalog are
inspected by eye, in order to remove spurious detections.  These
objects have to be removed because their shape measurements are
affected by cosmetic defects (such as dead columns, bleeding stars,
halos, diffraction spikes) or because the objects are likely to be
part of a resolved galaxy (e.g., HII regions). The visual inspection
is important as it is not possible to remove all spurious detections
in a fully automatic process. Their removal is crucial for an accurate
measurement of cosmic shear, because they increase the measurement of
the variance, and introduce artificial ellipticity correlations.

\vbox{
\begin{center}
\leavevmode
\hbox{%
\epsfxsize=8.0cm
\epsffile{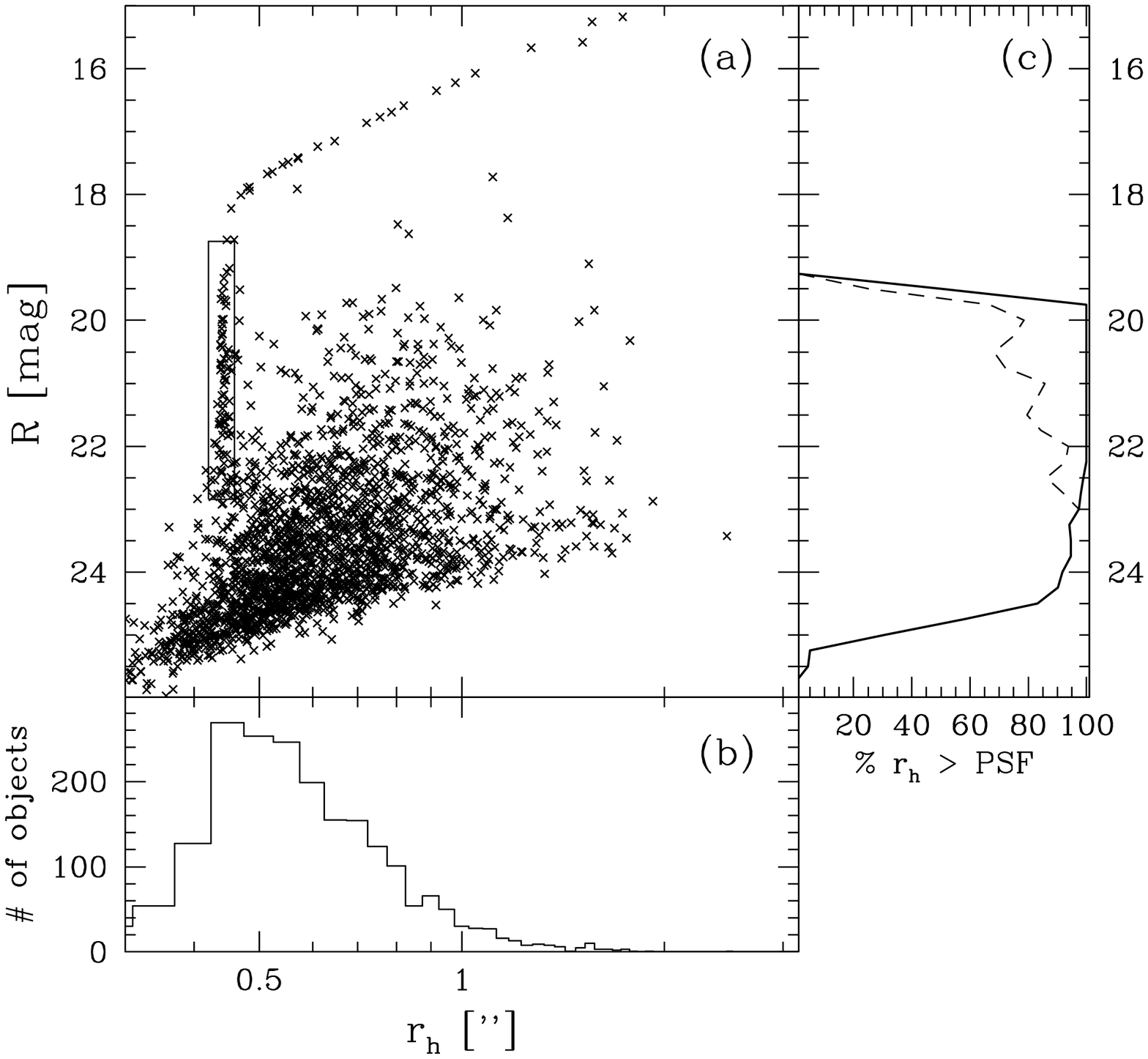}}
\figcaption{\footnotesize (a) Plot of the apparent $R_C$-band
magnitude versus the half-light radius. The vertical sequence of
points at $r_h\sim 0\farcs{45}$ (indicated by the rectangle) consists
of bright, unsaturated stars, that are used to study the PSF. (b)
Histogram of the number of objects of given $r_h$. (c) The dashed line
shows the fraction of objects with half light radii larger than the
PSF.  The solid line shows the fraction of objects larger than the
PSF, when the total counts are corrected for the contribution by
stars.  Objects larger than the PSF are assumed to be galaxies and
only these are used in the weak lensing analysis. The figure
demonstrates that down to $R\sim 24$ this separation selects most
galaxies, as almost all objects are larger than the PSF.
\label{sizemag}}
\vspace{-1cm}
\end{center}}

\begin{figure*}[b!]
\begin{center}
\leavevmode
\hbox{%
\epsfysize=8cm
\epsffile[20 165 520 560]{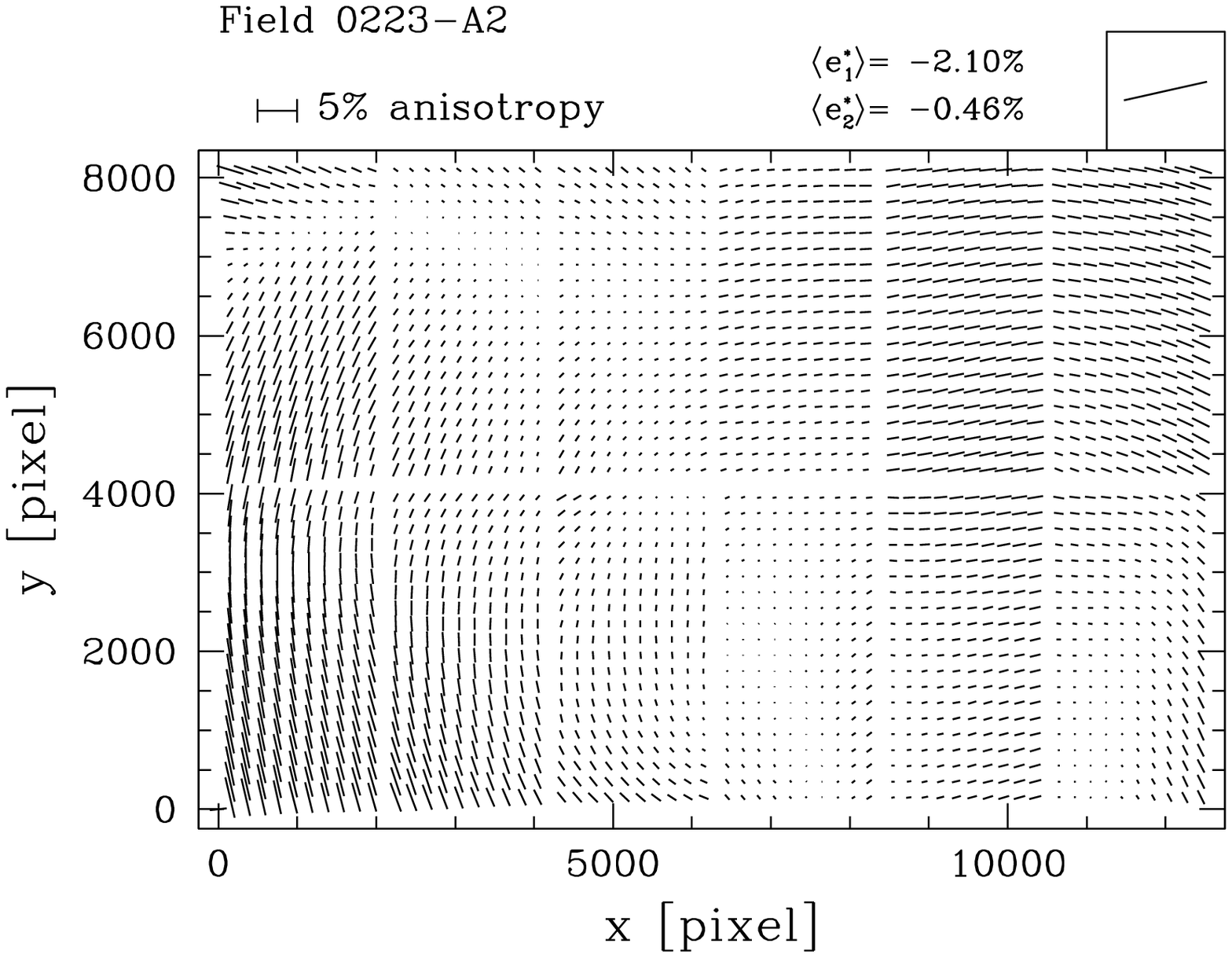}
\epsfysize=8cm
\epsffile[20 170 520 690]{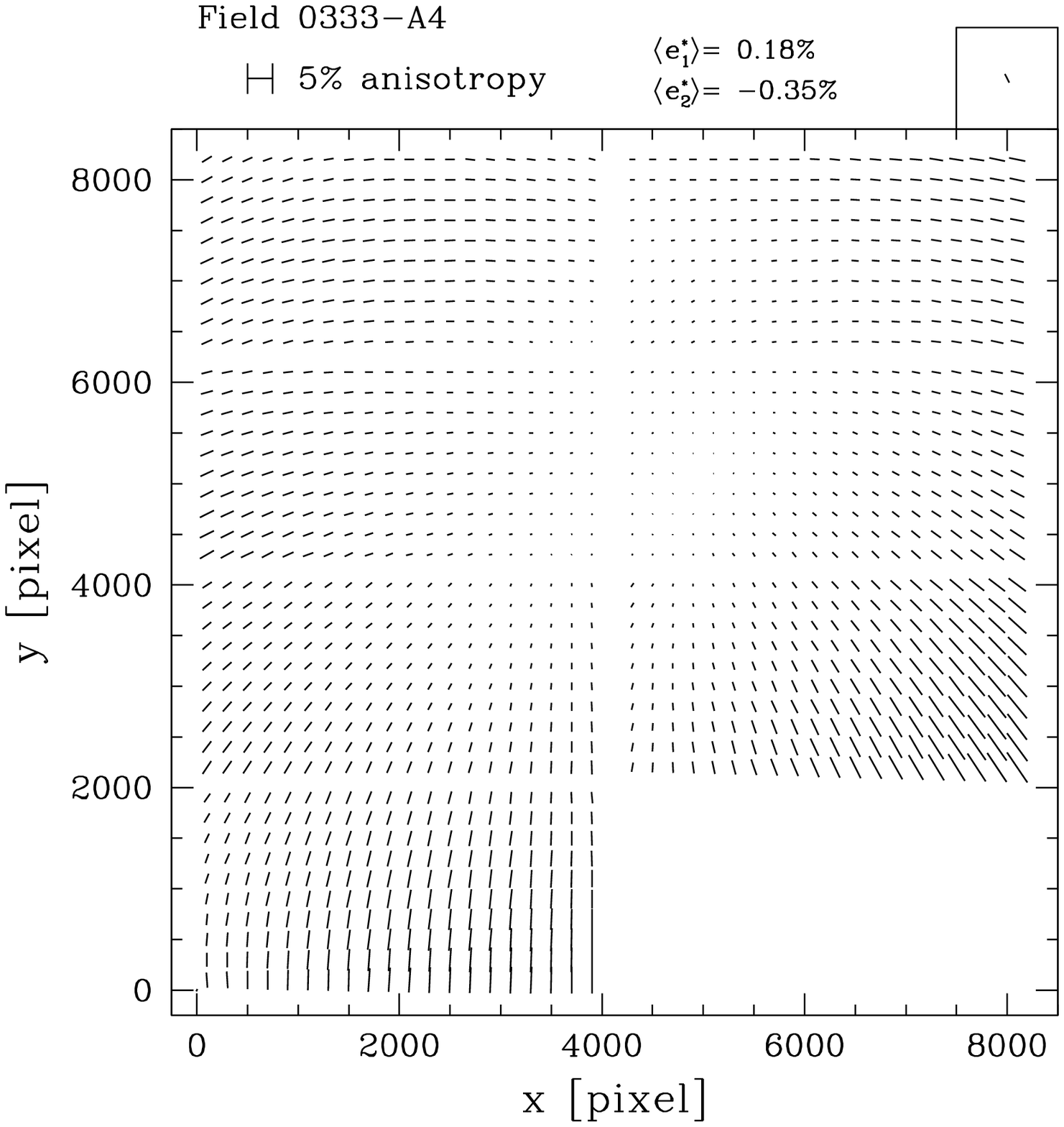}}
\begin{footnotesize}
\figcaption{Left panel: PSF anisotropy as a fuction of
position for one of the CFHT pointings; right panel: PSF anisotropy
for one of the CTIO pointings. One of the chips in the CTIO
observations was not functioning and has been omitted. The sticks
indicate the direction of the major axis of the PSF, and the length is
proportional to the observed ellipticity of the PSF. In order to show
the higher order spatial dependence of the anisotropy we have
subtracted the average ellipticity.  The direction of the average PSF
anisotropy is indicated in the top right box, and the amplitude is
indicated as well. Although the PSF anisotropy was determined from
fits to the observed shapes for individual chips, the figure clearly
shows a large scale dependence.
\label{psfan}}
\end{footnotesize}
\end{center}
\end{figure*}

\subsection{Correction for the PSF}

To measure the small, lensing induced distortions in the images
of the faint galaxies it is important to accurately correct the
shapes for observational effects, such as PSF anisotropy and seeing;
PSF anisotropy can mimic a cosmic shear signal, and a correction
for the seeing is required to relate the measured shapes to the 
real lensing signal.

To do so, we follow the procedure outlined in Hoekstra et al. (1998).
We select a sample of moderately bright stars from our observations,
and use these to characterize the PSF anisotropy and seeing.
Figure~\ref{sizemag}a shows a plot of the apparent $R_C$-band
magnitudes of the detected objects versus their measured half-light
radii for one of the chips of the A2 pointing of the 0223 patch
(seeing $\sim 0\farcs{77}$). We have also indicated the region from
which we have taken the stars used for the analysis of the PSF.

We fit a second order polynomial to the shape parameters of the
selected stars for each chip. This procedure is repeated for various
dispersions of the weight function (for details see Hoekstra et
al. 1998).  In the left panel of Figure~\ref{psfan} we present the
resulting model PSF anisotropy for the A2 pointing of the 0223
patch. To show in more detail the higher order spatial dependence of
the anisotropy we have subtracted the average ellipticity. Although
the fits were obtained from the individual chips, the mosaic image in
Figure~\ref{psfan} shows continuity between the chips.

The results for one of the CTIO pointings is presented in the right
panel of Figure~\ref{psfan}. Comparison of the patterns presented in
Figure~\ref{psfan} with other pointings shows that the pattern is
fairly stable, although the amplitute varies, because of focus
variations. In general the PSF anisotropy is small, a point which we
will address in more detail below, when we examine the residuals left
after correction of the shapes of the galaxies used in the weak
lensing analysis.

\subsection{Telescope distortion}

The effect of the PSF is not the only observational distortion that
has to be corrected. The optics of the camera stretches the
images of galaxies (i.e., it introduces a shear) because of the
non-linear remapping of the sky onto the CCD.  We have used
observations of astrometric fields to find the mapping between the sky
and the CCD pixel coordinate system, and derived the corresponding
camera shear.

\begin{figure*}
\begin{center}
\leavevmode
\hbox{%
\epsfysize=7.5cm
\epsffile[20 175 560 560]{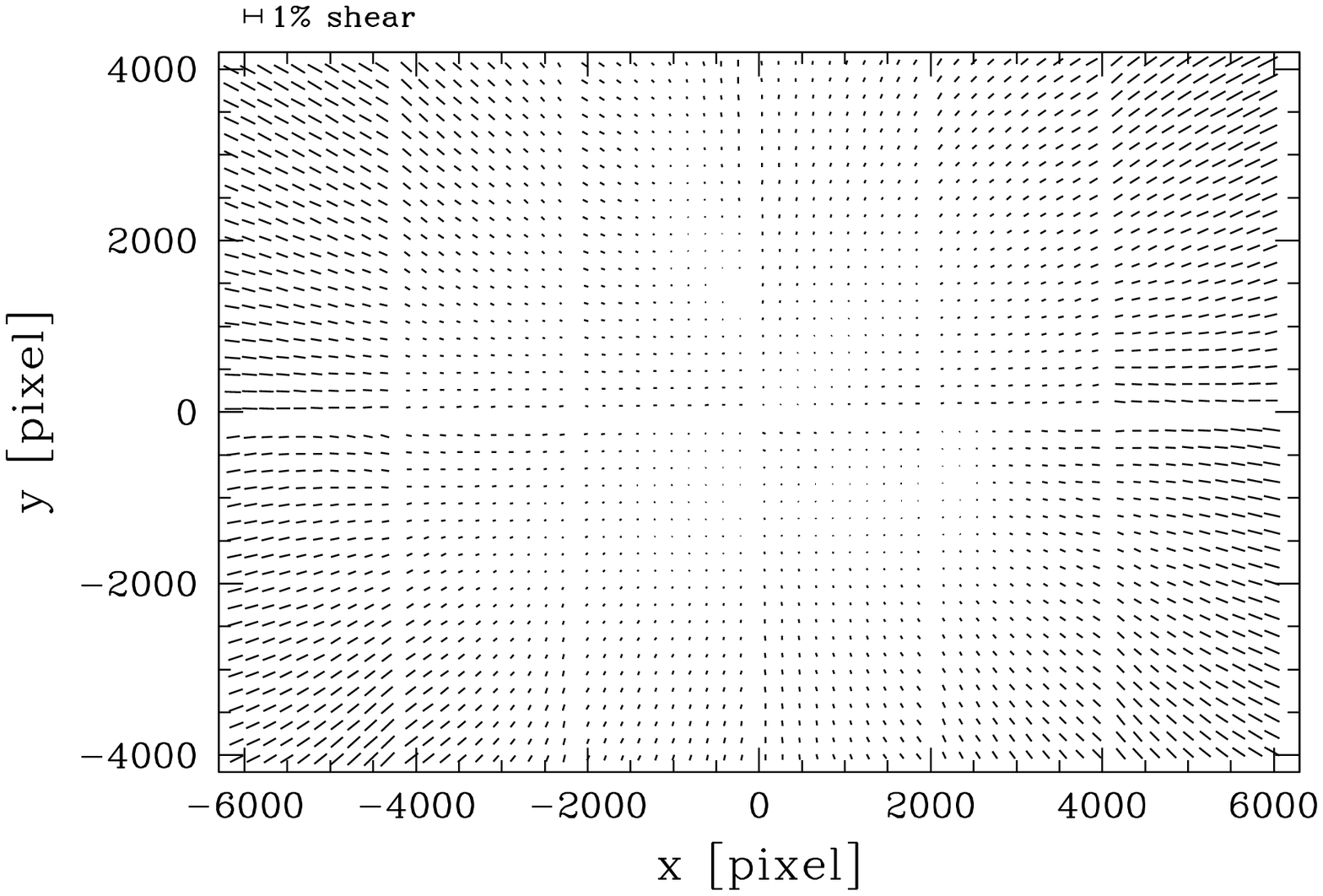}
\epsfysize=7.5cm
\epsffile{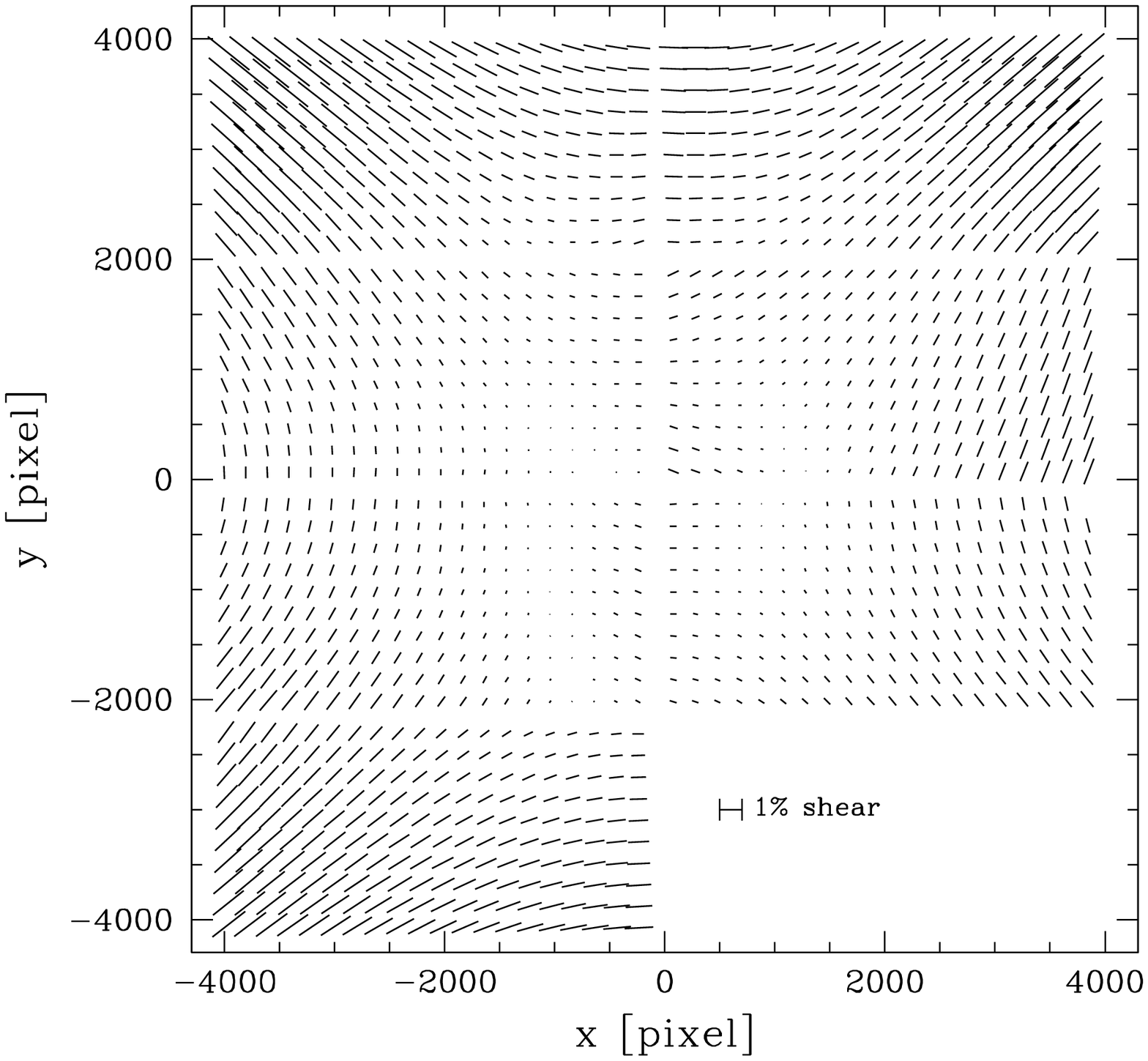}}
\figcaption{\footnotesize Left panel: camera distortion measured for
the CFH12k camera for run~2. The shear introduced by the camera is
small, reaching a maximum value of $\sim 1\%$ at the edges of the
field of view; Right panel: the camera distortion measured from the
CTIO data. One of the chips was not functioning and has been
omitted. The induced shear by this camera is somewhat larger than the
CFH12k camera, but still small: $\sim 2\%$ at the edges of the field
of view.
\label{camshear}}
\end{center}
\vspace{-0.5cm}
\end{figure*}

The camera shear for the CFH12k camera for run~2 is presented
in Figure~\ref{camshear}. The shear introduced by the camera is small,
reaching a maximum value of $\sim 1\%$ at the edges of the field
of view.

Other weak lensing studies, which use dithered observations, need to
remap the images before combining the data, thus removing the camera
distortion.  We have analysed single exposures, and measured the
shapes of the galaxies on the reduced images that have not been
remapped to remove the camera distortion. As discussed in Hoekstra et
al. (1998), the images of both the stars and galaxies are sheared by
the camera.  The measured shape of the PSF (as shown in
Figure~\ref{psfan}) is then the combination of PSF anisotropy and
camera shear, and therefore the real PSF anisotropy is somewhat
smaller than the measured value. The change in the ellipticity of an
object caused by the camera shear depends on its shape (and hence the
correction) whereas the PSF anisotropy correction depends mainly on
the size of the object. As a result the correction for PSF anisotropy
leaves a residual ellipticity, because the correction will be too
large for larger galaxies.  However, Hoekstra et al.  (1998)
demonstrated that the correction for the residual camera shear is
straightforward: one just needs to subtract the camera shear from the
measured shear, which is what we have done.

The camera shear is more or less radial with respect to the center of
the camera (although the camera shear of the CTIO Mosaic II camera
shows a significant non-radial component), which results in a negative
tangential shear. It is therefore useful to examine the average
tangential shear of the galaxies with respect to the center of the
camera. The results are presented in Figure~\ref{rem_cam}a. The solid
circles indicate the average tangential distortion of the galaxies
with respect to the center of the CFH12k camera after correction for
PSF anisotropy.  These measurements agree well with inferred camera
shear (solid line).  After subtracting the camera shear we obtain the
open circles, which are consistent with no signal.

\subsection{Residuals}

The correction scheme developed by Kaiser et al. (1995) has been
tested extensively (e.g., Hoekstra et al. 1998; Bacon et al.  2001;
Erben et al. 2001). The assumptions that have to made to derive the
original Kaiser et al. (1995) correction parameters do not necessarily
hold in real data, the modifications suggested by Hoekstra et al. (1998)
allow it to be applied to more complicated PSFs. This is supported 
by numerous simulations which indicate the method works remarkably well
down to the required accuracy for current data sets.

In addition we have tested the method using a realistic simulation.
The simulated data sets were created using the software tools {\tt
SkyMaker} and {\tt Stuff}\footnote{see {\tt
http://terapix.iap.fr/soft}}, which have been described in detail in
Erben et al. (2000). The simulated galaxies have realistic profiles,
with a mix of early type, late type galaxies, and disk/bulge ratios
matched to actual observations. The PSF is computed using realistic
pupil functions, and includes all the problems encountered in real
data, such as coma, tracking errors, aberration, spider arms from the
support of the secondary mirror, etc.

The simulation is described by Van Waerbeke et al. (in preparation),
and here we briefly discuss the results. Van Waerbeke used an N-body
simulation to infer the corresponding (cosmic) shear by means of
ray-tracing.

\vbox{
\begin{center}
\leavevmode
\hbox{%
\epsfxsize=8cm
\epsffile{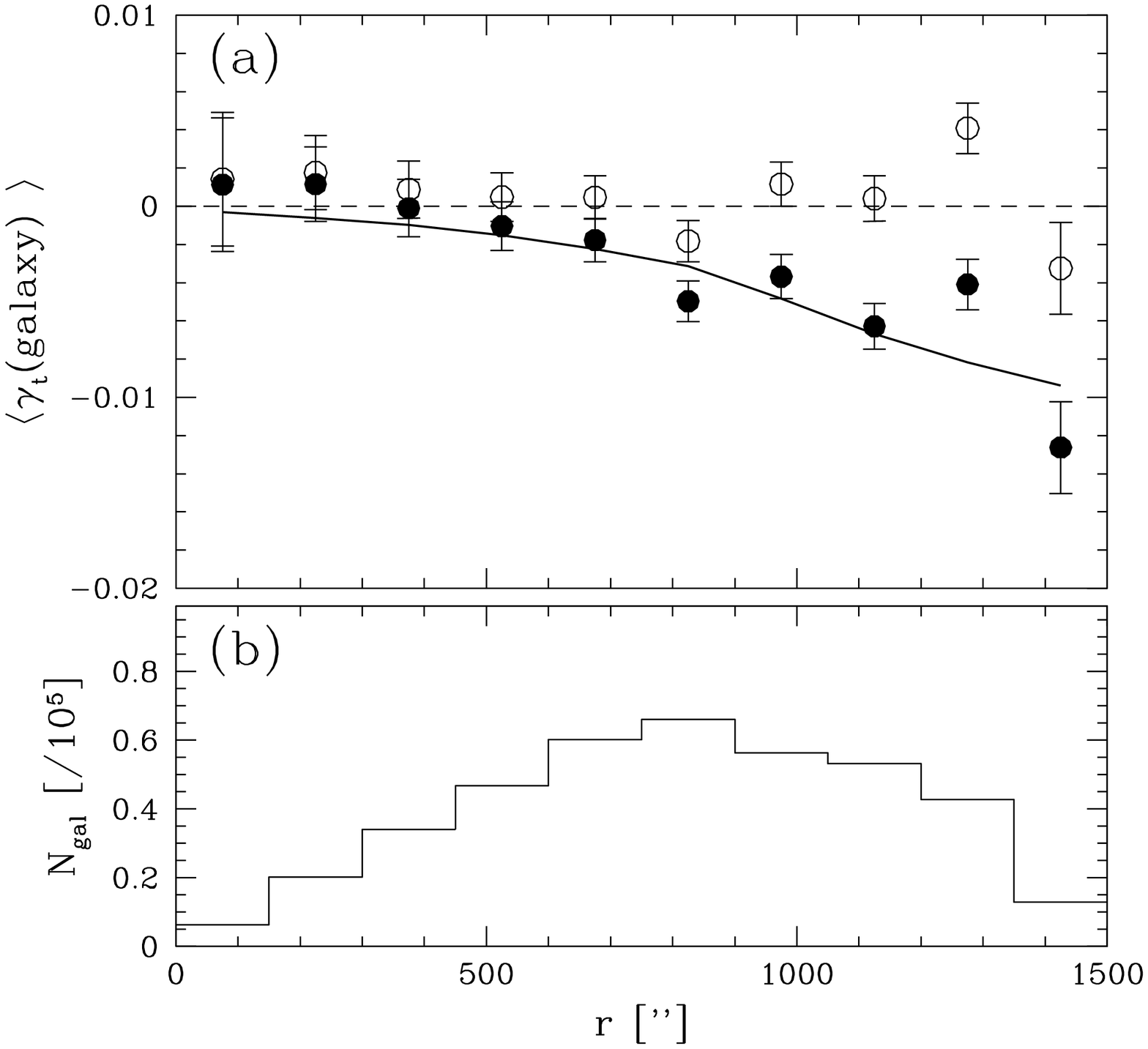}}
\figcaption{\footnotesize (a) Solid circles indicate the average
tangential distortion of the galaxies with respect to the center of
the CFH12k camera after correction for PSF anisotropy. The solid line
corresponds to the average tangential camera shear. The open circles
give the measurements of the galaxies after correcting for both PSF
anisotropy and camera distortion. The results indicate that the two
steps in the correction have worked well. (b) number of galaxies as a
function of radius used to produce panel~(a).
\label{rem_cam}}
\vspace{-0.5cm}
\end{center}}

\begin{figure*}
\begin{center}
\leavevmode
\hbox{%
\epsfxsize=7.6cm
\epsffile[20 150 470 700]{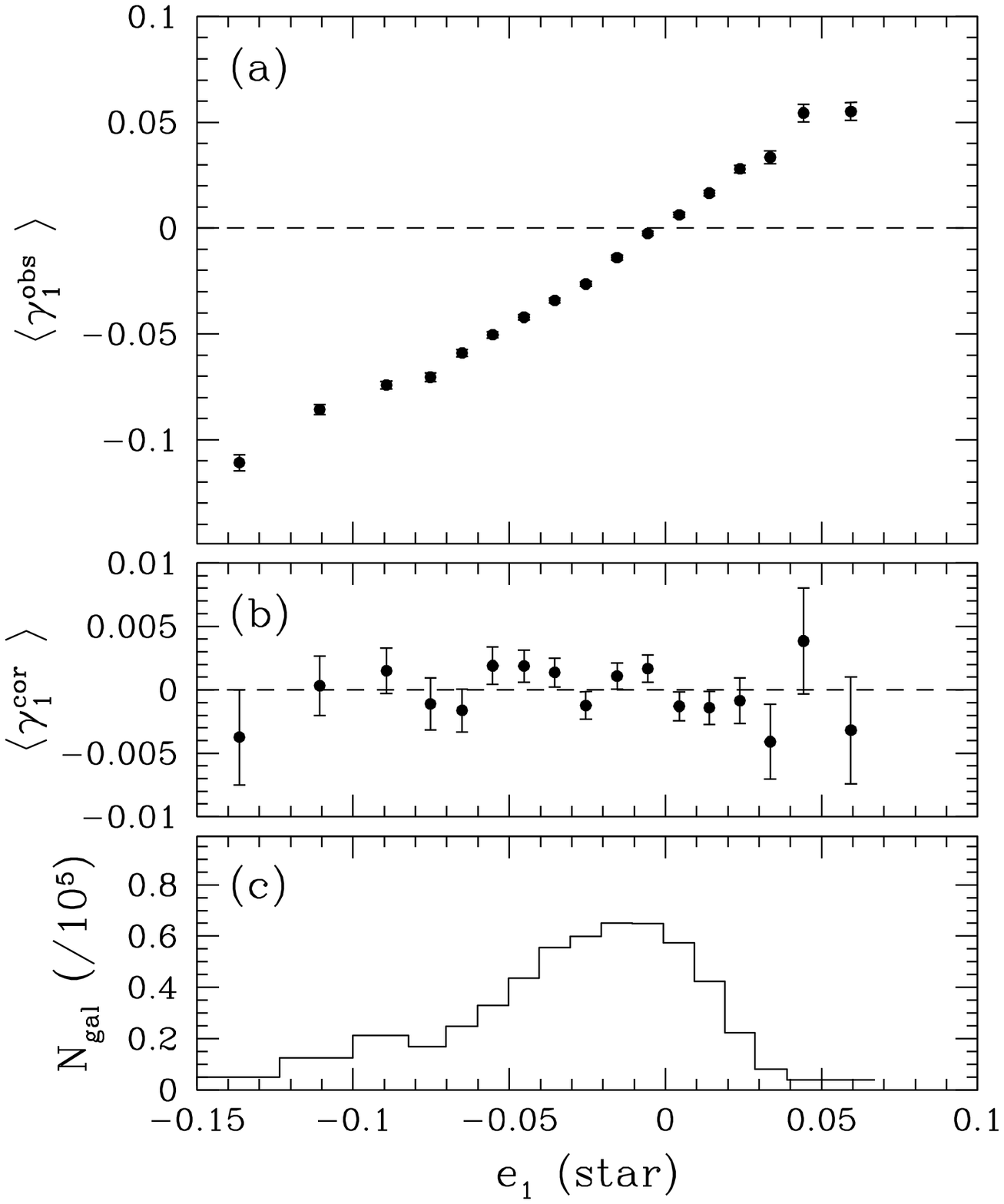}
\epsfxsize=7.6cm
\epsffile[20 150 470 700]{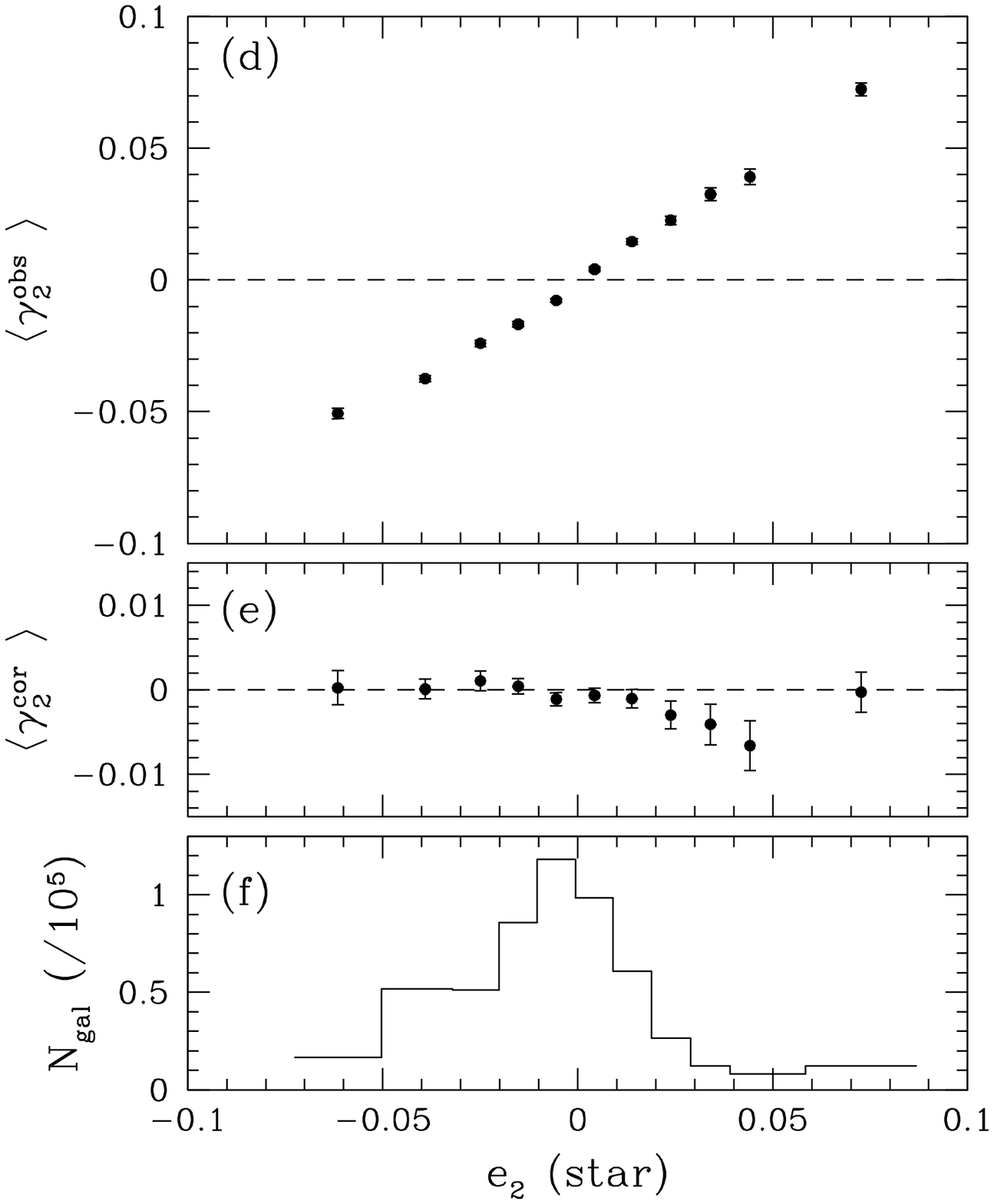}}
\begin{small}
\figcaption{\footnotesize (a) $\langle\gamma_1\rangle$ for the galaxies with
$20<R<24$ as a function of the value of the ellipticity component
$e_1$ of the stars used to correct the galaxies. The results show the
expected strong correlation; (b) The average $\gamma_1$ after
correction for PSF anisotropy (note the vertical scale has been
expanded with respect to panel~a); (c) The number of galaxies with a
given value for $e_1$ of the PSF; (d)-(f) same as (a)-(c), but for
$\gamma_2$. The residuals are consistent with no signal, demonstrating 
that the correction for the PSF anisotropy has worked well.
\label{psf_res}}
\end{small}
\end{center}
\end{figure*}

Artifical images of galaxies were sheared using these
results. These galaxies were ``observed'' and the final images are
convolved with a realistic PSF.  A mosaic of 900 images (corresponding
to $\sim 12$ deg$^2$), each with a different PSF was given to us. We
note that the PSFs used in the simulation had worse systematics than the
once observed in the RCS data.

We analysed these images in the same way as we have analysed the RCS
data, and measured the cosmic shear signal.  In doing so, we were able
to recover the input lensing signal, which was unknown to us. We were
able to recover the signal within 10\% ($\sim 1\sigma$) of the input
value. Because of the noise introduced by the intrinsic shapes of the
galaxies, larger simulations are required to test whether we can
measure the lensing signal to even higher accuracies.

We have also examined the residuals in the RCS data after PSF
correction. A useful test, although not definitive, is to plot the
average shape of the galaxies as a function of the shape of the
PSF. The results for the two components of the shear are presented in
Figure~\ref{psf_res}. Panels~(c), and~(f) indicate that the PSF
anisotropy is small for most galaxies.  The residuals presented
in panels~(b), and~(e) (note the different vertical scale) are consistent
with no signal. The results of Figure~\ref{psf_res} and the results from
the simulation suggest that we are able to measure the lensing signal to 
an accuracy better than 10\%.

\section{Cosmic shear signal}

To study the weak lensing signal caused by large scale structure, we
use the top-hat smoothed variance of the shear (Bacon et
al. 2000; Kaiser et al. 2000; Maoli et al. 2001; van Waerbeke et
al. 2000; van Waerbeke et al. 2001). Other statistics, such as the
ellipticity correlation function (Kaiser 1992; van Waerbeke
2001; Wittman et al. 2000) or the aperture mass statistic (Schneider
1998, Schneider et al. 1998; van Waerbeke 2001) have also been
used. 

Here we briefly discuss how the lensing signal depends on the
assumed cosmology and the redshift distribution of the sources.
Detailed discussions on this subject can be found elsewhere (e.g.,
Schneider et al. 1998; Bartelmann \& Schneider 2001).

Given a cosmological model, the variance in the shear caused by
large scale structure can be computed as a function of aperture
size $\theta$ (e.g., Jain \& Seljak 1997):

\begin{equation}
\langle\gamma^2\rangle(\theta)= 2\pi\int_0^\infty
dl~l~P_\kappa(l)\left[\frac{J_1(l\theta)}{\pi l \theta}\right]^2,
\end{equation}

\noindent where $\theta$ is the radius of the aperture used to compute
the variance, and $J_1$ is the first Bessel function of the first
kind. The effective convergence power spectrum  $P_\kappa(l)$ is
given by

\begin{equation}
P_\kappa(l)=\frac{9 H_0^4 \Omega_m^2}{4 c^4}
\int_0^{w_H} dw \left(\frac{\bar W(w)}{a(w)}\right)^2
P_\delta\left(\frac{l}{f_K(w)};w\right).
\end{equation}

\noindent Here $w$ is the radial coordinate, $w_H$ corresponds to the
horizon, $a(w)$ the cosmic scale factor, and $f_K(w)$ the comoving
angular diameter distance. As shown by Jain \& Seljak (1997) and
Schneider et al. (1998) it is necessary to use the non-linear power
spectrum in equation~(2). This power spectrum is derived from the
linear power spectrum following the prescriptions from Peacock \&
Dodds (1996).

$\bar W(w)$ is the source-averaged ratio of angular diameter distances
$D_{ls}/D_{s}$ for a redshift distribution of sources $p_w(w)$:

\begin{equation}
\bar W(w)=\int_w^{w_H} dw' p_w(w')\frac{f_K(w'-w)}{f_K(w')}.
\end{equation}

Thus it is important to know the redshift distribution of the sources,
in order to relate the observed signal to $P_\kappa(l)$. A detailed
discussion of the adopted redshift distribution can be found in
Section 4.1. Figure~\ref{gg_mag} shows the top-hat smoothed variance
$\langle\gamma^2\rangle$ on a scale of 1 arcminute as a function of
limiting magnitude of the sample of sources. To compute the signal we
used galaxies with $20<R<R_{\rm lim}$ and used the photometric
redshift distribution inferred from the Hubble Deep Fields north and
south (see section 4.1). The top axis indicates the corresponding
median redshift of the source galaxies.

One of the advantages of deep observations is obvious from
Figure~\ref{gg_mag}: the signal increases quickly for limiting
magnitudes fainter than $R=24$ (or higher median redshift). In
addition the number of sources increases as well, resulting in higher
signal-to-noise ratios of the measurements.

\subsection{Redshift distribution}

In order to relate the observed lensing signal to physical parameters,
such as $\sigma_8$ or $\Omega_m$, knowledge of the redshift
distribution of the sources is crucial. The galaxies used in weak
lensing surveys are generally too faint to be included in redshift
surveys, and little is known about their redshift distribution from
spectroscopic studies.

Compared to the other, deeper, cosmic shear studies, our analysis has
the major advantage that the redshift distribution of the sources we
use is better known. Down to a limiting magnitude of $R_C=24$ the redshift 
distribution has been determined spectroscopically by Cohen et al. (2000),
although this survey is limited to a relatively small patch of sky,
and is likely to suffer from cosmic variance. 

In addition the galaxies are larger, which is demonstrated in
Figure~\ref{sizemag}: down to $R_C=24$ the galaxies are easily
separated from the stars. This has the advantage over deeper surveys
(where the fainter galaxies have sizes comparable to the PSF) in that
selecting objects larger than the PSF does not change the redshift
distribution significantly. Even for the worst seeing images
considered here $(\sim 1\farcs{1})$ the stars are well separated from
the galaxies for $R_C=24$.

Photometric redshift studies, in particular those based on the Hubble
Deep Fields (e.g., Fern{\'a}ndez-Soto et al. 1999; Chen et al. 1998)
have also provided important information. The results of Hoekstra et
al. (2000) have demonstrated that they generally work well. However,
Hoekstra et al. (2000) noted a difference between the redshift
distributions inferred for the Northern and the Southern field, and
such field to field variation is not unexpected. However, currently
little is known about the amplitude of such variations, and more
studies are required to constrain the redshift distributions of these
faint galaxies.

\vbox{
\begin{center}
\leavevmode
\hbox{%
\epsfxsize=8cm
\epsffile{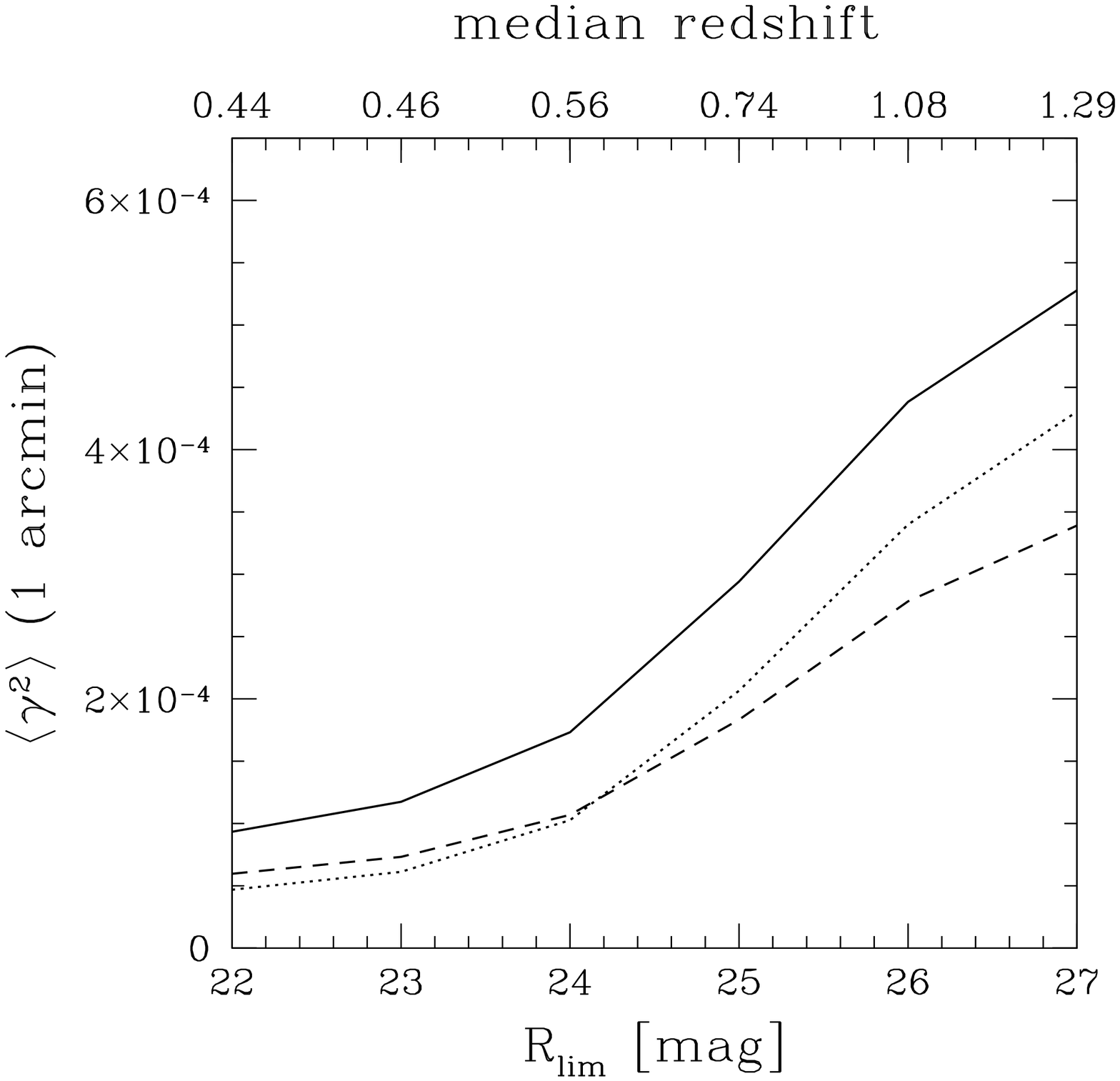}}
\figcaption{\footnotesize The expected variance induced by large scale
structure in an aperture of radius 1 arcminute, as a function of
limiting magnitude of the source galaxies. To derive these results we
used source galaxies fainter than $R=20$, and assumed perfect shape
measurements. The result for the SCDM (solid line, $\Omega_m=1,~
\Omega_\Lambda=0,~\sigma_8=0.5,~\Gamma=0.5$), OCDM (dashed line,
$\Omega_m=0.3,~ \Omega_\Lambda=0,~\sigma_8=0.85,~\Gamma=0.21$) and
$\Lambda$CDM (dotted line,
$\Omega_m=0.3,~\Omega_\Lambda=0.7,~\sigma_8=0.9,~\Gamma=0.21$) are
shown. The cosmic shear signal increases rapidly with increasing
limiting magnitude, or increasing median redshift (as indicated by the
top axis).
\label{gg_mag}}
\end{center}}

In order to minimize the contribution of cosmic variance to the
redshift distribution, we use the photometric redshift distributions
from Fern{\'a}ndez-Soto et al. (1999) and Chen et al. (1998) to
compute the predicted lensing signal for a given cosmology (see Fig.~2
from Hoekstra 2001 for the resulting redshift distribution). To do so,
we have to take into account that the uncertainty in the shape
measurements depend on the apparent magnitudes (and thus on the
redshifts) of the sources: the contribution of distant, small faint
galaxies (with noisy shape measurements) to the measured lensing
signal is smaller compared to brighter galaxies.

This is illustrated in Figure~\ref{weight}. Figure~\ref{weight}a shows
the expected (based on modeling of deep number counts), the observed
(galaxies for which shapes could be measured), and effective number
counts (dotted line) as a function of apparent magnitude. The
effective number density takes into account the uncertainty in the
galaxy shapes, and gives a good indication of which galaxies
contribute most to the measurement of the lensing signal. Based on the
results displayed in Figure~\ref{weight}, we decided to use galaxies
with $20<R_C<24$ for the lensing analysis.

The relative weight (normalised to unity for bright galaxies) as a
function of magnitude is shown in Figure~\ref{weight}b. This weight
function is simply the inverse square of the uncertainty in the shape
measurement (see Hoekstra et al. 2000 for details), and reflects the
fact that the shapes estimates of faint galaxies are more noisy. We
derive the ``effective'' redshift distribution using this weight
function.  This effective redshift distribution is used to compute the
predicted lensing signals discussed in Section~5.

\vbox{
\begin{center}
\leavevmode 
\hbox{%
\epsfxsize=7.8cm 
\epsffile[15 160 580 705]{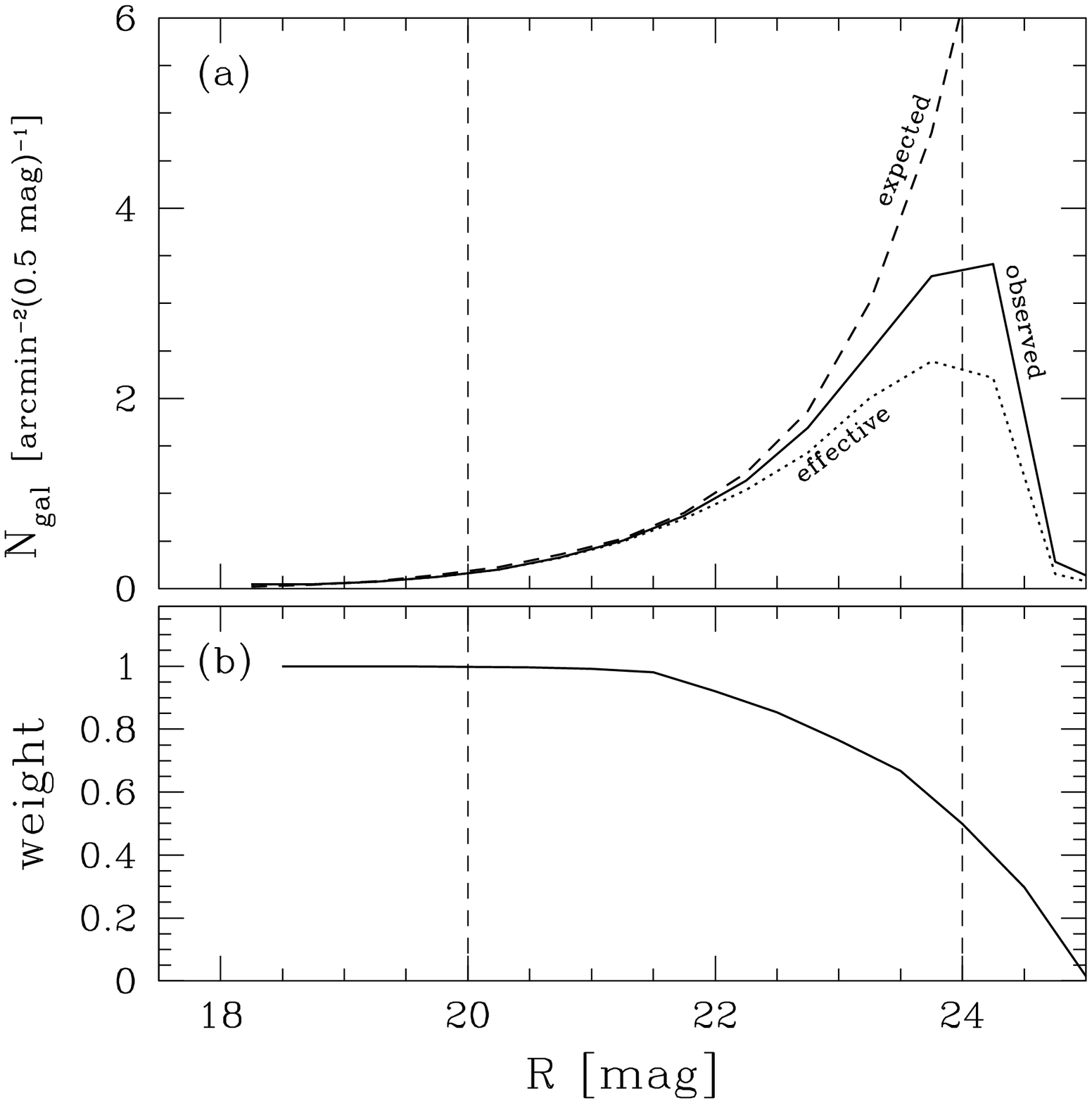}} 
\figcaption{\footnotesize (a) Observed number density of galaxies
(i.e. galaxies for which shapes could be determined) as a function of
apparent $R_C$ magnitude (solid line). The expected number density
(based on modeling of observed number counts) is indicated by the
dashed line. (b) The weight (based on the uncertainty in the shape
measurements; see Hoekstra et al. 2000) multiplied by the completeness
fraction as a function of apparent magnitude. The product of the
number of galaxies and the weight gives a good indication of the
relative contribution to the lensing signal (dashed line in panel
(a). The result shows that most of the signal comes from galaxies
around $R_C=23.5$. The vertical dashed lines indicate the region
$20<R_C<24$, the range of apparent magnitudes for the source galaxies
we will use in the lensing analysis.
\label{weight}}
\vspace{-0.2cm}
\end{center}}

\section{Results}

\subsection{Observed signal}

In this section we present the measurement of the weak lensing signal
caused by large scale structure using the top-hat smoothed variance of
the shear. This statistic has been used by other groups to detect the
cosmic shear signal (e.g., Bacon et al. 2000; Kaiser et al.  2000;
Maoli et al. 2001; Van Waerbeke et al. 2000, 2001).  The top-hat
smoothed variance is fairly insensitive to errors in the analysis
because residual shears are added in quadrature.  Consequently the
observed signal can always be considered as an upper limit, because
residual errors always increase the variance. However, the results
presented in Figure~\ref{psf_res} and the simulations discussed in
section~3.3 indicate that we can measure the shapes of the galaxies
accurately.

As was found by van Waerbeke et al. (2000), close pairs of galaxies can
introduce an excess signal, because of overlapping isophotes.  We
therefore remove pairs with a separation of less than $2\farcs16$
(which corresponds to 10 pixels for the CFHT data, and 8 pixels for
the CTIO data). This lowered the signal at small scales ($\sim$ 20\%
for an aperture of radius 1 arcminute). We note that on these scales
intrinsic alignments can also be important. To compute the top-hat
smoothed variance we use the practical estimators given in Van
Waerbeke et al. (2001).

Figure~\ref{compscat} shows the top-hat smoothed variance as a
function of scale for both the CFHT data (filled circles) and the CTIO
data (open circles) using galaxies with $20<R_C<24$. The errorbars are
estimated from a large number of random realisations of the data set
where the orientations of the galaxies were randomized.

Note that the measurements at various scales are strongly correlated,
and this causes all the CTIO measurements at large scales to be higher
than the CFHT results. The results obtained from the two different
telescopes agree very well with one another. Because the systematics
for the two data sets are different, this excellent agreement
demonstrates that the various observational biases  have been removed
successfully.

Another useful test is to compare the signals from the individual
patches. Figure~\ref{var_150} displays the top-hat smoothed variance
of the shear for an aperture of radius 2.5 arcminutes (where the
signal-to-noise ratio is highest) for the 13 observed patches. The
measurements for the individual patches agree with the ensemble
average.

It is also important to examine the correction for the circularization
by the PSF, because it determines the amplitude of the signal. To do
so, we computed the variance on a scale of 2.5 arcminutes for each
pointing, and looked for a correlation with seeing. The results show
no trend with seeing.

Figure~\ref{var_all} shows the measurement of the top-hat smoothed
variance for the full weak lensing data set . The signal-to-noise
ratio of our measurements is very good, reaching a maximum of $\sim 6$
at a radius of 2.5 arcminutes.

For comparison the predictions for three different
cosmological models are also shown in this figure. All three models
are good fits to the data, indicating the need for additional observational
constraints (cf. \S 5.2)

\vbox{
\begin{center}
\leavevmode
\hbox{%
\epsfxsize=8.5cm
\epsffile[44 155 580 705]{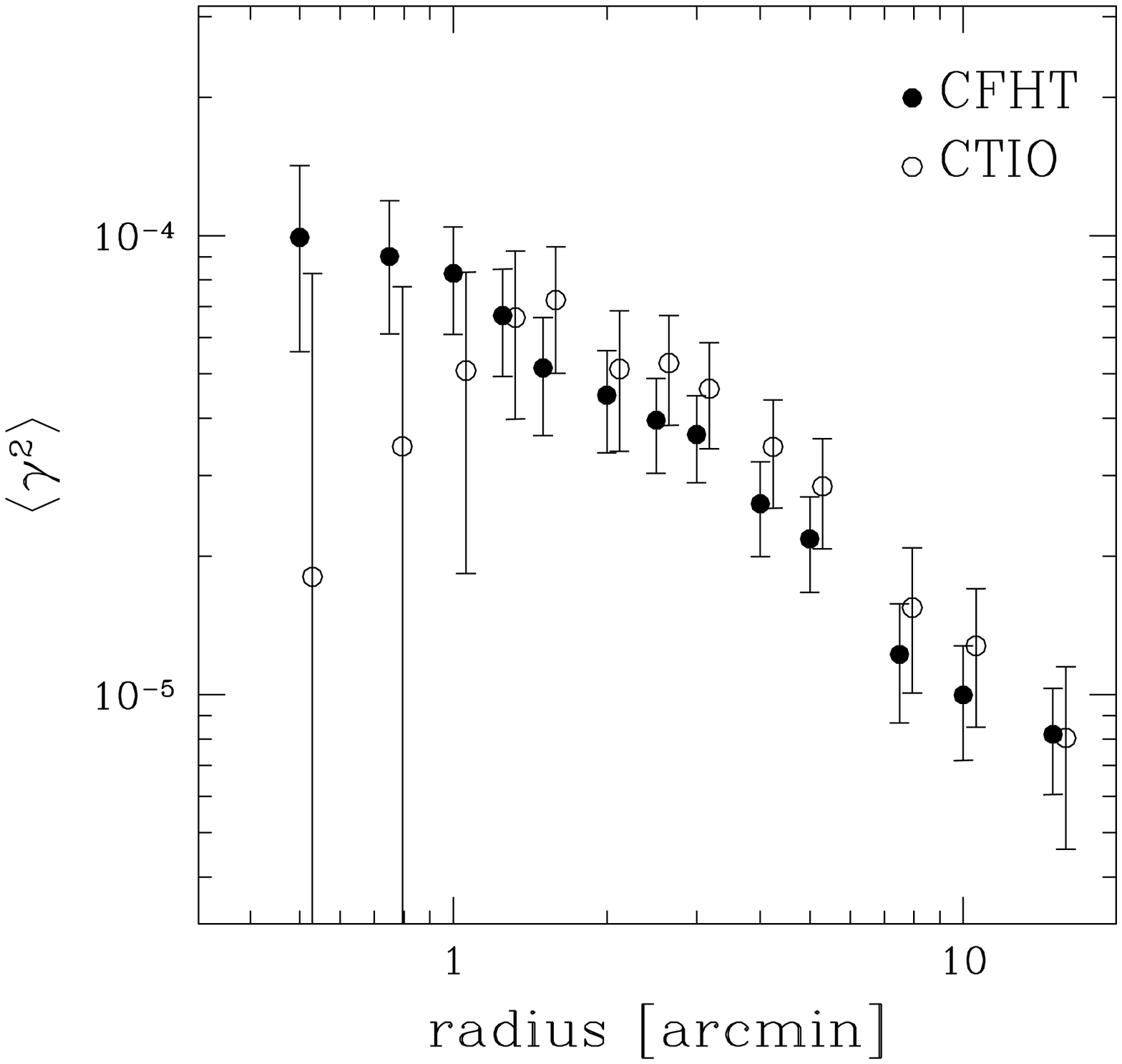}}
\figcaption{\footnotesize Top-hat smoothed variance of the shear as a
function of aperture radius. The filled circles indicate the
measurements based on 16.4 deg$^2$ of CFHT data, and the open circles
correspond to the analysis of 7.6 deg$^2$ of CTIO data. For display
purposes the CTIO points have been offset slightly in radius.  Note
that the measurements at various scales are strongly correlated, and
this causes all the CTIO measurements at large scale to be higher than
the CFHT results. The results obtained from the two different
telescopes agree well with one another.
\label{compscat}}
\end{center}}

\vbox{
\begin{center}
\leavevmode
\hbox{%
\epsfxsize=8.5cm
\epsffile[15 250 580 710]{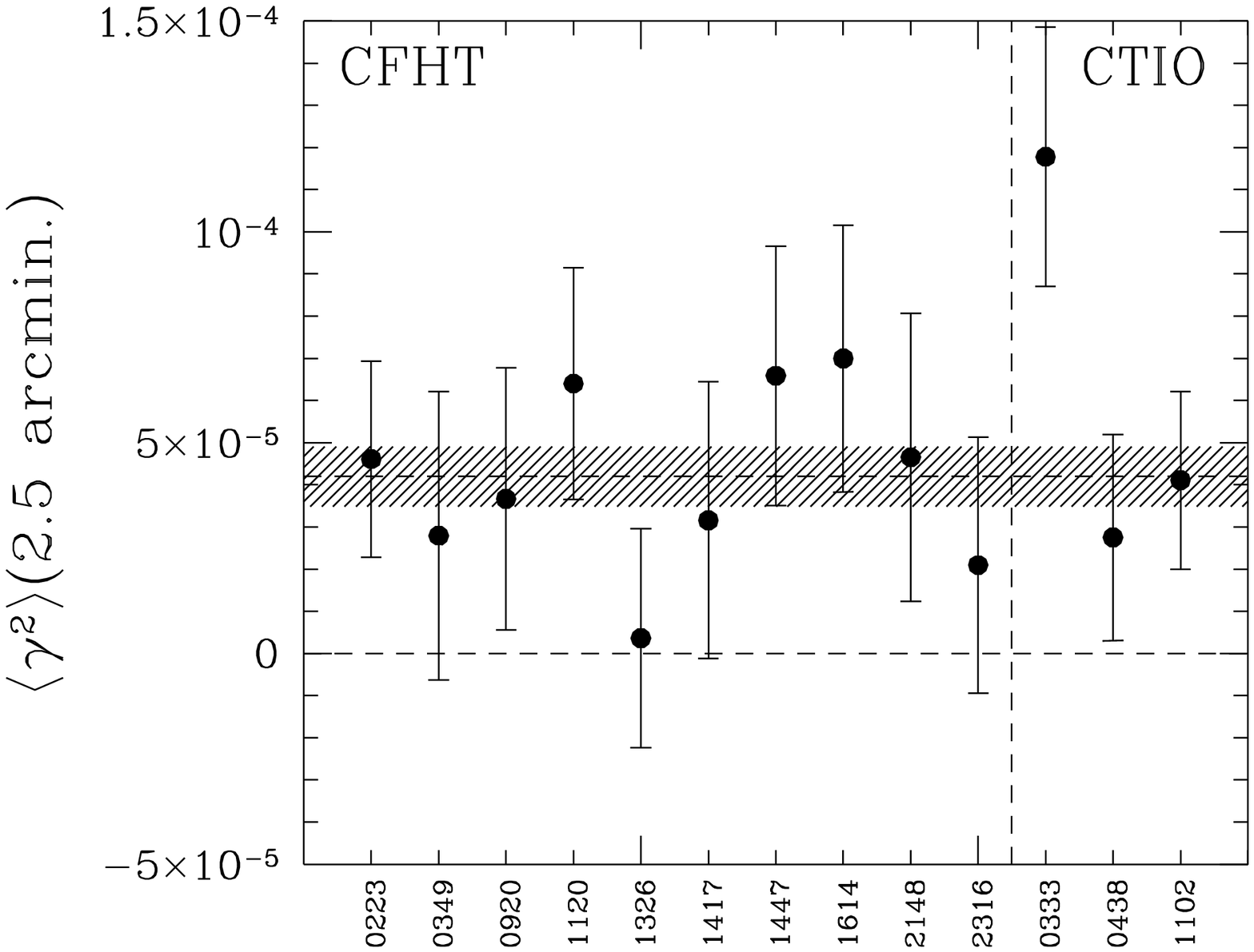}}
\figcaption{\footnotesize The top-hat smoothed variance of the shear for an
aperture of radius 2.5 arcminutes for the 13 observed patches.
The shaded region corresponds to the $1\sigma$ limits around
the average. The measurements for the individual patches
agree well with the average, indicating that the cosmic 
variance is small.
\label{var_150}}
\end{center}}

It is difficult to compare our measurements directly to most other
studies, because of the difference in the source redshift distribution
(caused by the diffence in filters and integration times). However, we
can compare directly to the results from Bacon et al. (2000), who have
used a similar cut in apparent magnitude. Bacon et al. (2000) find a
variance of $\langle\gamma^2\rangle=(24\pm7)\times 10^{-5}$ in
$8\times 8$ arcminute cells. This scale is similar to a radius of
$\sim 4$ arcminutes for which we find
$\langle\gamma^2\rangle=(2.8\pm0.5)\times 10^{-5}$. Our signal is much
lower than their result. Bacon et al. (2000) note a residual
correlation between the shape of the PSF and the galaxies (their
Fig.~7), and this might explain their increased variance. We note that
because of the large errorbar on the measurement of Bacon et
al. (2000) the results are consistent at the $3\sigma$ level.

\subsection{Constraints on cosmological parameters}

As described in Section~4 the predicted amplitude of the top-hat
smoothed variance depends on the various cosmological parameters,
and therefore provides a powerful method to constrain these parameters. 
Unfortunately several degeneracies exist (e.g., Bernardeau et
al. 1997; Jain \& Seljak 1997). These studies show that the amplitude
of the signal is mainly determined by a combination of $\sigma_8$ and
$\Omega_m$, although the shape parameter $\Gamma$ ($\sim \Omega_m h$ 
in a CDM cosmology) is also important.

As demonstrated by Van Waerbeke et al. (2001), it is possible to
partially break the degeneracy between $\sigma_8$ and $\Omega_m$ if
priors on the shape of the power spectrum are assumed.  The value of
$\Omega_\Lambda$ can be constrained by combining the measurements of
the lensing signal for sources at different redshifts, using the fact
that the angular diameter distances depend on $\Omega_\Lambda$. The
combination of the weak lensing measurements and the constraints from
studies of the fluctuations of the cosmic microwave background (CMB)
will provide much stronger constraints of the parameters, as the
combination will break the degeneracies.

\vbox{
\begin{center}
\leavevmode
\hbox{%
\epsfxsize=8cm
\epsffile[40 180 580 705]{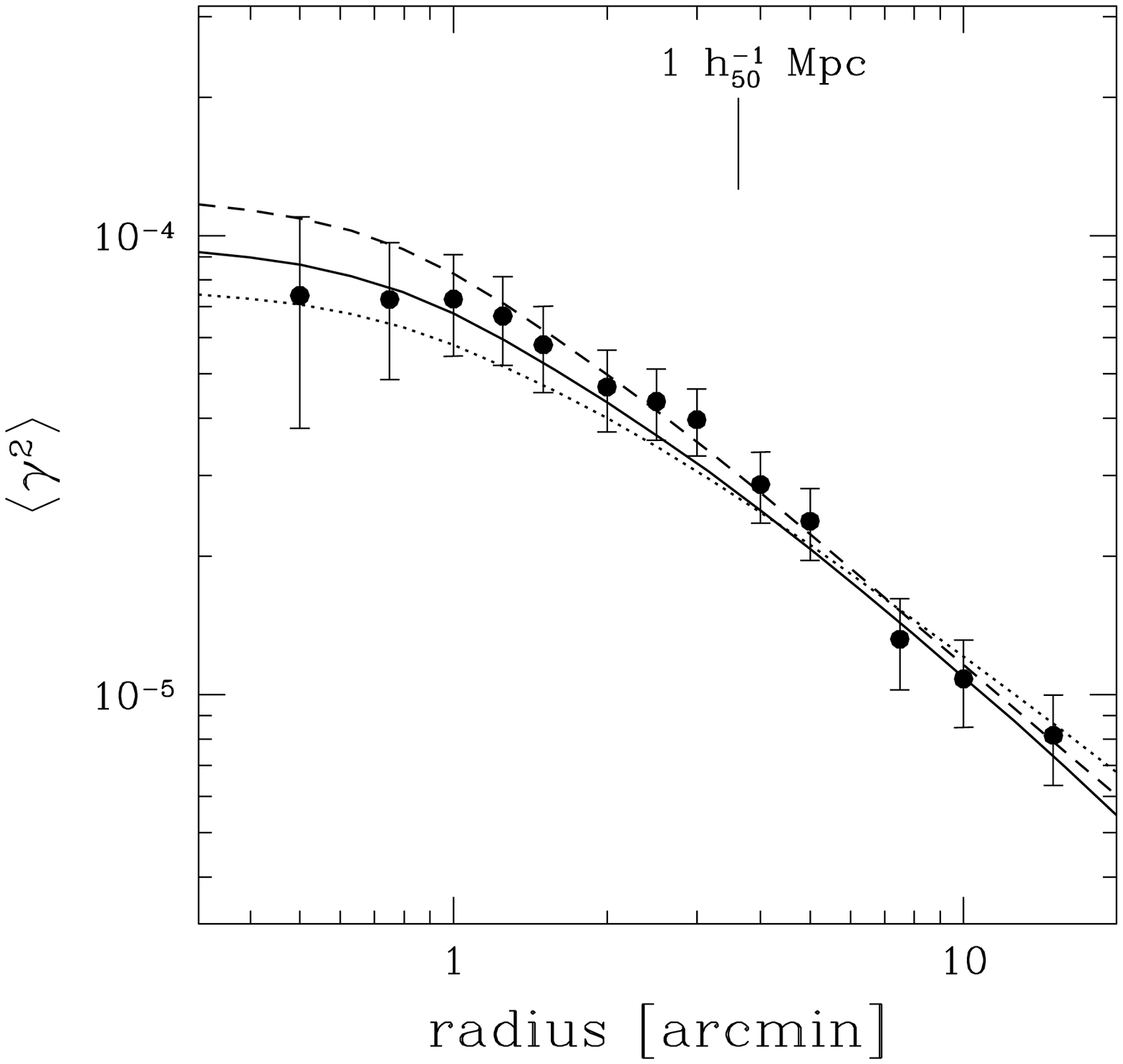}}
\figcaption{\footnotesize Measurement of the top-hat smoothed variance
(excess variance caused by lensing by large scale structure) using
galaxies with $20<R_C<24$.  The data consist of 16.4 deg$^2$ of CFHT
data and 7.6 deg$^2$ of CTIO data. The drawn lines correspond to the
best fit SCDM
($\Omega_m=1$,~$\Omega_\Lambda=0,~\Gamma=0.7,~\sigma_8=0.31$; solid
line), the best fit OCDM
($\Omega_m=0.3$,~$\Omega_\Lambda=0.0,~\Gamma=0.21,~\sigma_8=0.86$;
dashed line), and the best fit $\Lambda$CDM
($\Omega_m=0.3$,~$\Omega_\Lambda=0.7,~\Gamma=0.21,~\sigma_8=0.81$;
(dotted line) model. We fixed $h=0.7$, which gives a high value for
$\Gamma$ for the SCDM model. Without additional constraints on the
cosmological parameters, the lensing results are consistent with a
wide range of cosmological models. The errorbars are estimated from a
large number of random realisations of the data set where the
orientations of the galaxies were randomized.  Note that the points at
various scales are strongly correlated.  Under the assumption that the
lensing structures are halfway between the observer and the sources,
we have indicated a scale of $1~h_{50}^{-1}$ Mpc.
\label{var_all}}
\end{center}}

Here we will use the measurements of the top-hat smoothed variance to
find constraints on $\Omega_m$ and $\sigma_8$. We assume that the
measurements follow a normal distribution. We computed the covariance
matrix from a large number of random realisations of the data,
thus including the survey geometry in the noise correlation.

We compute the model predictions using equation~3, under the
assumption that $\Gamma=0.21$, and using the effective redshift
distribution discussed in section~4.1. We use this value for $\Gamma$
to allow for a direct comparison with the results from Van Waerbeke et
al. (2001).  The predictions are compared to the observations, and the
likelihood for the combination of $\Omega_m$ and $\sigma_8$ is
computed. The results for models with $\Omega_\Lambda=0$ are presented
in Figure~\ref{cont_OCDM}. The contours indicate the 68.3\%, 95.4\%,
and 99.7\% confidence limits on two parameters jointly.

The results for the $\Omega_m+\Omega_\Lambda=1$ model are presented in
Figure~\ref{cont_LCDM}. The best fit value for $\Omega_m$ is lower
than for the OCDM model, but no strong constraints on $\Omega_m$ and
$\sigma_8$ can be placed. Allowing for larger values ($\Gamma \sim
0.7$), the goodness of fit for high $\Omega_m$ models is comparable to
those of low $\Omega_m$ models. We note, however, that studies of the
galaxy correlation function suggest values for $\Gamma=0.1-0.3$
(e.g., Dodelson et al. 2001)

The best fit $\Lambda$CDM cosmology yields
$\sigma_8=0.81^{+0.14}_{-0.19}$ (95\% confidence; $\Gamma=0.21$). For
the best fit OCDM model we obtain $\sigma_8=0.86^{+0.14}_{-0.17}$
(95\% confidence). Van Waerbeke et al. (2001) find
$\sigma_8=0.99^{+0.08}_{-0.10}$ (95\% confidence) for an OCDM model,
which is in fair agreement with our result, in particular when the
uncertainty in the redshift distribution used by Van Waerbeke et
al. (2001) is taken into account.

To investigate the agreement between the CFHT and CTIO data we
also computed the best fit values for $\sigma_8$ for the two
data sets. We obtain $\sigma_8=0.84\pm0.09$ (68\% confidence) for
the CFHT data, and $\sigma_8=0.76\pm0.12$ (68\% confidence) for
the CTIO data. The values agree well, and the probability of
finding a smaller difference is 40\%.

\vbox{
\vspace{-0.2cm}
\begin{center}
\leavevmode
\hbox{%
\epsfxsize=8cm
\epsffile{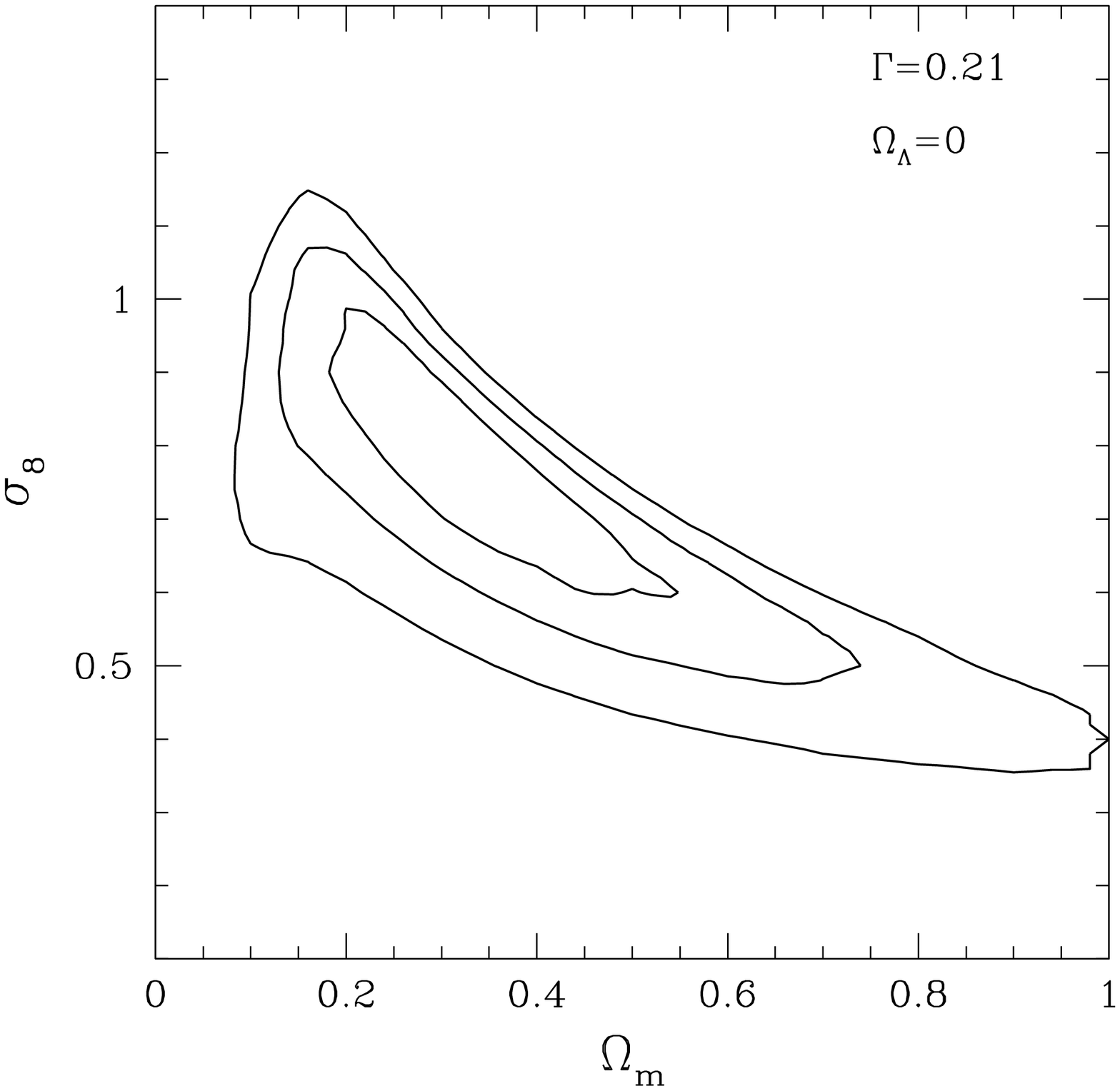}}
\vspace{-0.2cm}
\figcaption{\footnotesize Likelihood contours as a function of
$\Omega_m$ and $\sigma_8$, inferred from the analysis of the top-hat
smoothed variance. We have used only the measurements at radii $\ge$ 1
arcminute. The contours have been computed by comparing the
measurements to CDM models with $n=1$, $\Gamma=0.21$ and
$\Omega_\Lambda=0$.  The contours indicate the 68.3\%, 95.4\%, and
99.7\% confidence limits on two parameters jointly.
\label{cont_OCDM}}
\end{center}}

\vbox{
\begin{center}
\leavevmode
\hbox{%
\epsfxsize=8cm
\epsffile{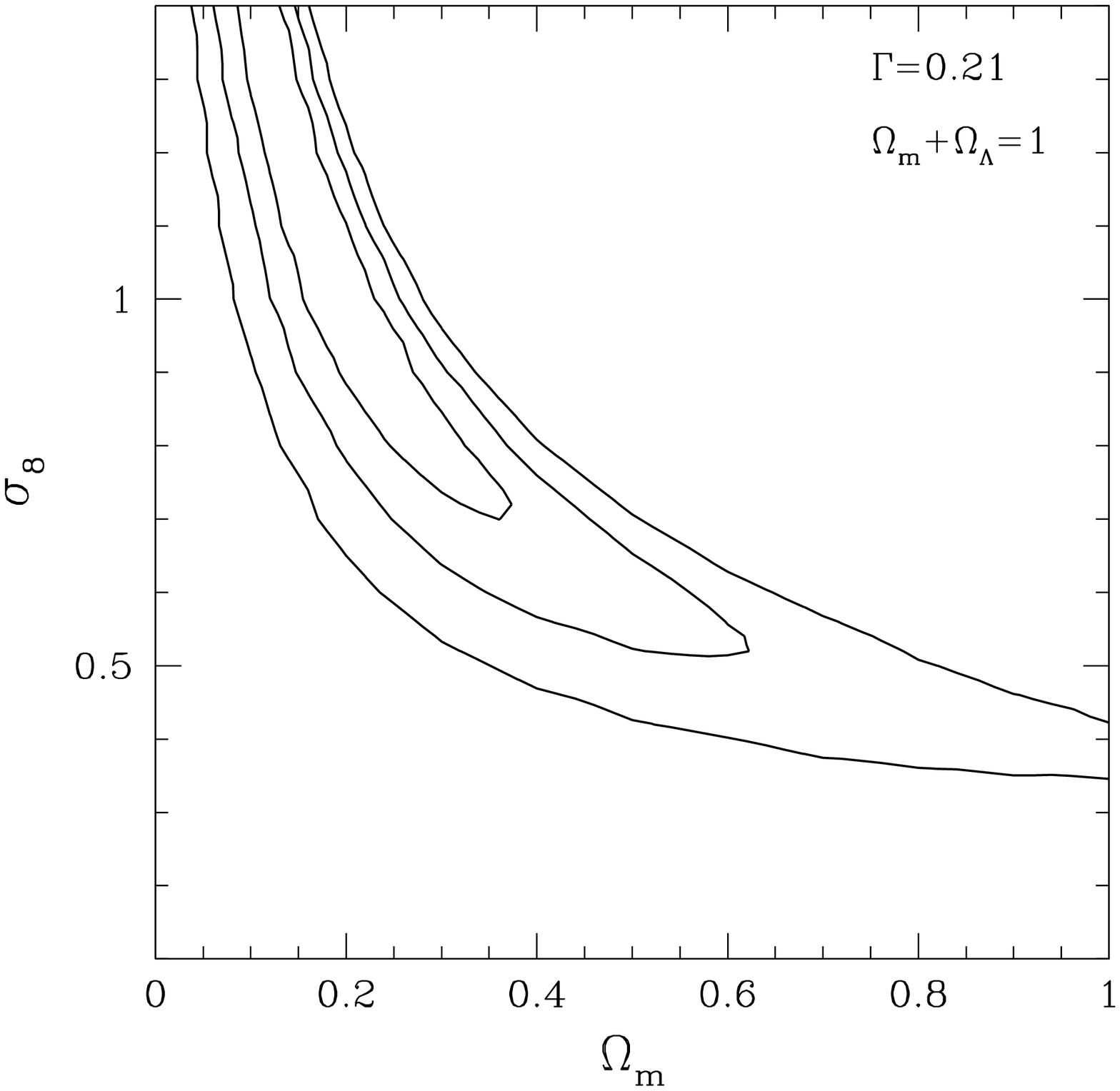}}
\figcaption{Same as Figure~\ref{cont_LCDM}, but now for an
$\Omega_m+\Omega_\Lambda=1$ CDM cosmology. 
\label{cont_LCDM}}
\vspace{-0.2cm}
\end{center}}

The RCS probes a different redshift range than the study presented by
Van Waerbeke et al. (2001). Although both results are based on the
same correction scheme, there are many differences in the various
steps in the analyses. Thus the agreement found here suggests that
accurate measurements of cosmic shear can be made.

\section{Conclusions}

We have analysed $\sim 24$ square degrees of $R_C$-band imaging data
from the Red-Sequence Cluster Survey to study the weak lensing
signal caused by intervening large scale structure. To minimize the
effect of cosmic variance, the measurements have been obtained from 13
patches that are widely separated on the sky. We have used data from
two different telescopes: $\sim 16.4$ square degrees of CFHT data and
$\sim 7.6$ square degrees of CTIO data.  We detect the signal with high
confidence on scales ranging from 1 to 30 arcminutes using galaxies
with $20<R_C<24$.

Because of the various observational distortions which need to be
corrected for, a careful analysis of the residuals is important.  The
results suggest that we have successfully corrected for the
systematics. In addition we compared the measurements from CFHT and
CTIO and find excellent agreement.

Compared to other studies of cosmic shear, the RCS imaging data
is relatively shallow. This has the disadvantage that the lensing
signal is low. However, the galaxies are larger (which results in
smaller corrections for the PSF), and the redshift distribution
is known fairly well (which is important for determining cosmological
parameters). 

Intrinsic alignments of galaxies contaminate the lensing signal.  This
is particularly important for lensing studies that use low redshift
galaxies. The median redshift of our sample of sources is $\sim 0.5$,
and predictions of the contribution of the intrinsic alignments
suggest it is small for the results presented here. However, the
effect of intrinsic alignments is expected to be comparable to the
errorbars of the full RCS data set, and eventually needs to be
corrected for.  The RCS will be complemented with $B$ and $V$ band
imaging which will provide photometric redshifts for a large number of
galaxies.  With such a data set we will be able to measure the
intrinsic alignments by selecting galaxies at similar photometric
redshifts.

We use the photometric redshift distribution inferred from the Hubble
Deep Fields to relate the measured top-hat smoothed variance to
estimates of cosmological parameters. Because of degeneracies in the
parameters we can only place constraints on $\Omega_m$ and $\sigma_8$
jointly. For the currently favoured $\Lambda$CDM model
$(\Omega_m=0.3,~\Omega_\Lambda=0.7,~\Gamma=0.21)$ we obtain
$\sigma_8=0.81^{+0.14}_{-0.19}$ (95\% confidence), in good agreement
with the results from Van Waerbeke et al. (2001).

The RCS data and the observations used by Van Waerbeke are quite
different, and also the weak lensing analyses are somewhat
different. Thus the agreement found here suggests that accurate
measurements of cosmic shear can be made.

\acknowledgments It is a pleasure to thank Ludo van Waerbeke for many
discussions, and providing a useful test of the correction scheme used
in this paper. We would like to thank the CFHT and CTIO TACs for their
generous allocation of telescope time. This research is partially
supported via an operating grant from NSERC to HKCY. MDG acknowledges
support from NSERC via PGSA and PGSB scholarships. PBH acknowledges
financial support from Chilean grant FONDECYT/1010981. LFB's research
is funded by FONDECYT, Chile under Proyecto 1000537.

\end{document}